\documentclass[prb,twocolumn,grouppedaddress]{revtex4-1}
\usepackage{amssymb}
\usepackage{graphicx}
\usepackage{hyperref}
\usepackage{amsfonts}
\usepackage{amsmath,amssymb,amsfonts,amsthm}
\usepackage{bm}
\usepackage{color}
\usepackage{slashed}
\usepackage{braket}

\def\be{\begin{equation}}
\def\ee{\end{equation}}
\def\bea{\begin{eqnarray}}
\def\eea{\end{eqnarray}}
\def\ba{\begin{array}}
\def\ea{\end{array}}

\begin{document}

\title{Collective excitations of a system of coupled relativistic and non-relativistic two-dimensional electron gases}

\author{Ajit C. Balram}
\email{ajit@phys.psu.edu}
\author{Jimmy A. Hutasoit}
\author{Jainendra K. Jain}
\affiliation{Department of Physics, The Pennsylvania State University, University Park, PA 16802}

\begin{abstract}
We study collective excitations of a two-dimensional electron system consisting of two kinds of charge carriers: relativistic or Dirac electrons with linear dispersion and non-relativistic electrons with parabolic dispersion. We find that the strength of the interaction between the different charge carriers, \textit{i.e.}, the inter-species interaction, plays a significant role in determining the number of plasmon modes as well as their dispersions in certain parameter regimes.
\end{abstract}

\maketitle

\section{Introduction}
In recent years there has been a great deal of interest in the electronic properties of novel materials like graphene and topological insulators (TIs). A major reason is that unlike normal metals and insulators they host relativistic massless Dirac electrons with a linear dispersion (graphene in the bulk and TIs at their surface states). Considerable efforts have been made, both theoretically and experimentally, to understand the collective excitations of electrons in these materials. An example of one such excitation is the plasmon, namely the collective oscillation of electrons in a conducting medium. In graphene, both single layer and multi-layer, the relativistic plasmon dispersion has been known for quite some time \cite{hwang2007dielectric,PhysRevB.87.235418,PhysRevB.85.085443}. Ref. \onlinecite{Luo2013351} reviews the recent progress and applications of plasmons in graphene. For a TI, the Dirac plasmon dispersion was calculated by Refs. \onlinecite{PhysRevLett.104.116401,efimkin2012collective,
efimkin2012collective2,efimkin2012spin}. Very recently there has also been a report of observing the Dirac plasmons in a TI for the first time \cite{di2013observation,Okada2013_nature_nanotech}. Plasmon dispersion has also been calculated for thin-film TIs \cite{PhysRevB.85.085443} and there is a proposal to show interesting phenomena such as spin-charge separation of plasma oscillations in this system \cite{PhysRevB.88.205427}. 

In this article, we consider the collective excitations of a coupled system of 2-dimensional (2D) Dirac fermions and 2D non-relativistic electrons that coexist at the surface of a TI and interact via Coulomb interaction. Such a coexistence has been seen experimentally on the surface of Bi$_2$Se$_3$  \cite{bianchi2010coexistence} by angle-resolved photoemission spectroscopy (ARPES) and in BiSbTe$_3$ \cite{Xiang:2014fk} via magneto-transport measurements. These materials are promising candidates for three-dimensional TIs. Ref. \onlinecite{Okada2013_science} reports the observation of such a co-existence by scanning tunneling microscopy (STM) in a \emph{crystalline} TI, Pb$_{1-x}$Sn$_{x}$Se. We calculate the dispersion of the plasmon mode(s) within the Random Phase Approximation (RPA), paying special attention to the effect of the inter-species interaction. The difference of the interaction strength between the inter-species and intra-species interactions can come from, for example, the displacement between the two planes in which the two species are localized. 
The problem of the plasmon dispersion for a bilayer system has been considered in several articles \cite{Takada77,PhysRevB.23.805}. Plasmons for a non-relativistic two-dimensional electron system in proximity to a graphene sheet was considered by Principi {\em et al.} \cite{PhysRevB.86.085421}. That work did not consider, as appropriate for their system of interest, species changing part of the interaction (that is, tunneling between graphene and the GaAs quantum well), but shares some similarities with our results; in particular, they also find an approximately linear plasmon close to the edge of the single particle excitation spectrum region.

We find that when the strength of the inter-species interaction is of the same order as that of the intra-species one, there exists only a single plasmon mode. This is contrary to the naive expectation that there should exist two plasmon modes, one each corresponding to in-phase and out-of phase oscillations of the two species of electrons. To gain insight into the disappearance of one mode, we monitor the collective modes as a function of the inter-species interaction. When the inter-species interaction is completely switched off, there are obviously two plasmon modes. When the inter-species interaction is much smaller than the intra-species interaction, we find that for a small enough non-relativistic electron mass, there exist two plasmon modes. However, as the strength of the inter-species interaction increases, the plasmon mode with the lower energy enters into the single particle excitation region and gets damped. That is the reason why when the inter-species interaction is roughly equal to the intra-
species one, there exists only a single plasmon mode.

We also find that for the parameter regime relevant to the experiment \cite{bianchi2010coexistence}, the inter-species interaction plays a role in determining the long wavelength behavior of the plasmon mode. In particular, when the inter-species interaction is of equal strength to the intra-species one, we find that the plasmon mode exhibits a linear dispersion.

This article is organized as follows. In Section \ref{set}, we describe our set-up and calculate the plasmon dispersion using the RPA. The solution is obtained numerically and we show our results in Section \ref{res}. We end with a summary and discussions in Section \ref{summ}.

\section{Calculation of the plasmon dispersion within the RPA \label{set}}
Let us start by describing the two species of electronic excitations at the surface of a TI. The relativistic surface states are described by the wave function
\begin{eqnarray}
\ket{\psi_{\pm}({\bf p})}&=&\frac{1}{\sqrt{2}}
\begin{pmatrix} 
  e^{-i\phi_{{\bf p}}/2} \\
  \pm i e^{i\phi_{{\bf p}}/2}  
\end{pmatrix},
\end{eqnarray}
and their energy is given by
\be
\varepsilon_{\pm} = \pm v \, |p| - \mu,
\ee
with $\pm$ denoting the upper or lower Dirac cone, respectively, and $\phi_{\bf p}$ is the polar angle of the momentum ${\bf p}$. For non-relativistic electrons, we have
\begin{eqnarray}
 \ket{\psi_{\uparrow}({\bf p})}=\frac{e^{i{\bf p}.{\bf r}}}{\sqrt{2}} 
 \begin{pmatrix} 1 \\ 
		 0 
  \end{pmatrix}  \\ \nonumber
 \ket{\psi_{ \downarrow}({\bf p})}=\frac{e^{i{\bf p}.{\bf r}}}{\sqrt{2}} 
 \begin{pmatrix} 0 \\ 
		 1 
  \end{pmatrix}  \\ \nonumber
\end{eqnarray}
where $\uparrow$, $\downarrow$ correspond to spin up and down, respectively, and the energy is given by
\be
\varepsilon_{\rm nr}= \frac{p^2}{2m} + \Delta - \mu.
\ee

The interaction Hamiltonian is written as
\begin{widetext}
\bea
H_{\rm int} = \sum_{\mathbf{q,p,p'}} \sum_{i,j,m,n} \sum_{\gamma_i,\gamma_j,\gamma_m,\gamma_n} &&V_{im}^{jn}(\mathbf{q}) 
f_{\gamma_i \gamma_j}(\mathbf{p,p+q}) f_{\gamma_m \gamma_n}(\mathbf{p',p'-q})  a^{\dagger}_{\gamma_j}(\mathbf{p+q}) a^{\dagger}_{\gamma_n}(\mathbf{p'-q}) a_{\gamma_m}(\mathbf{p}') a_{\gamma_i}(\mathbf{p}),
\eea
\end{widetext}
where the $i,j,m,n$ are species indices and for Dirac fermions $i={\rm rel}$, $\gamma_i = \gamma_{\rm rel}=\pm$ correspond to conduction and valence bands, respectively, while for non-relativistic fermions $i = {\rm nr}$, $\gamma_i =\gamma_{\rm nr} = \uparrow$ and $\downarrow$ correspond to spin up and down, respectively.
The matrix elements are given by $f_{\gamma_i \gamma_j}(\mathbf{p,p+q}) = \langle \psi_{\gamma_j}({\mathbf{p+q}})|\psi_{\gamma_i}({\mathbf{p}}) \rangle$. 

\begin{figure}[t]
\begin{center} 
\includegraphics[width=3.5 in]{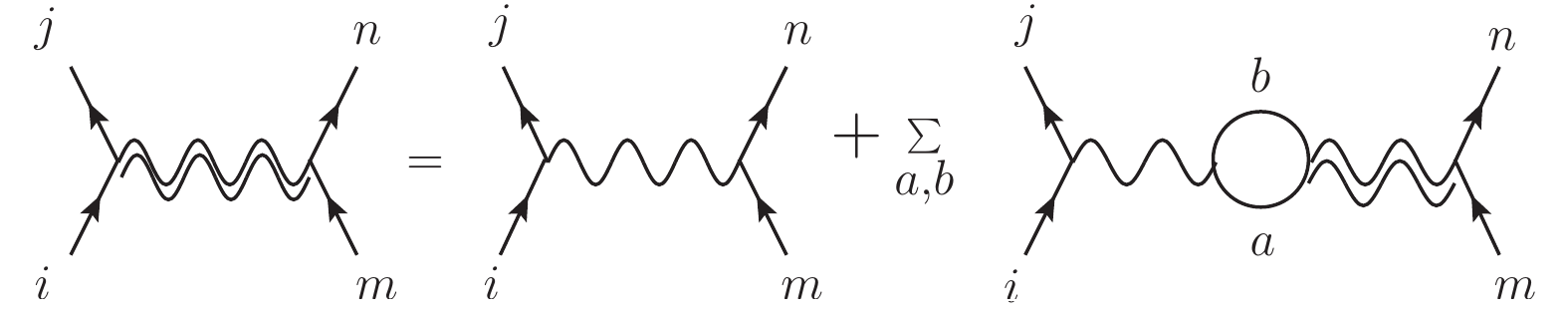}
\end{center}  
\caption{Feynman diagrams for the interaction. The doubly wiggly lines correspond to the full potential $\tilde{V}$, while the wiggly line corresponds to bare potential $V$. \label{fig:feyn1} }
\end{figure}

Within RPA, we have (see  Fig. \ref{fig:feyn1})
\bea
\tilde{V}_{i m}^{j n} = V_{i m}^{j n} + \sum_{a, b}V_{i a}^{jb} \Pi_{ab} \tilde{V}_{a m}^{bn}.
\eea
Therefore, 
\bea
\sum_{a, b} \left(\delta_{i a} \delta_{j b} - V_{i a}^{jb} \Pi_{ab} \right) \tilde{V}_{a m}^{bn} \equiv \sum_{a, b} {\cal E}_{i a}^{jb}  \tilde{V}_{a m}^{bn} = V_{i m}^{j n} \nonumber \\
\Rightarrow  \tilde{V}_{i m}^{j n} =\sum_{a, b} [{\cal E}^{-1}]_{i a}^{jb} {V}_{a m}^{bn}. \label{eq:tensor}
\eea
The singularities of ${\cal E}^{-1}$ correspond to the poles of the ``2-particle'' propagator, which are the plasmon modes. These can be obtained by solving 
\be
\det {\cal E}=0. \label{eq:det}
\ee
In the basis where $\{ \{i,j\}, \{a,b\} \} = \{\{\rm nr, nr\},\{\rm nr, rel\},\{\rm rel, nr\},\{\rm rel, rel\}\}$, we have
\begin{widetext}
\be
{\cal E} =    \begin{pmatrix} 
      1 - V_{\rm nr, nr}^{\rm nr, nr} \Pi_{\rm nr} & - V_{\rm nr, nr}^{\rm nr, rel}\Pi_{int} & -  V_{\rm nr, rel}^{\rm nr, nr}\Pi_{int} & - V_{\rm nr, rel}^{\rm nr, rel} \Pi_{\rm rel}\\
      -  V_{\rm nr, nr}^{\rm rel, nr} \Pi_{\rm nr} & 1 -  V_{\rm nr, nr}^{\rm rel, rel} \Pi_{\rm int} & - V_{\rm nr, rel}^{\rm rel, nr} \Pi_{\rm int} & - V_{\rm nr, rel}^{\rm rel, rel} \Pi_{\rm rel}\\
      -  V_{\rm rel, nr}^{\rm nr, nr}\Pi_{\rm nr} &  - V_{\rm rel, nr}^{\rm nr, rel}  \Pi_{\rm int} & 1 - V_{\rm rel, rel}^{\rm nr, nr} \Pi_{\rm int} & - V_{\rm rel, rel}^{\rm nr, rel} \Pi_{\rm rel}\\
       -  V_{\rm rel, nr}^{\rm rel, nr} \Pi_{\rm nr} & - V_{\rm rel, nr}^{\rm rel, rel}  \Pi_{\rm int} & - V_{\rm rel, rel}^{\rm rel, nr} \Pi_{\rm int} & 1 - V_{\rm rel, rel}^{\rm rel, rel}  \Pi_{\rm rel}
   \end{pmatrix},
\ee
\end{widetext}
where $\Pi_{\rm nr, rel, int}$ are the contributions from the non-relativistic electrons, Dirac electrons and the inter-species interaction to the polarization, respectively.

The contribution from the Dirac fermions is given by
\be
\Pi_{\rm rel} (\omega, q) = \sum_{\gamma, \gamma' = \pm}  \Pi_{\gamma \gamma'} (\omega, q), \label{eq:pirel}
\ee
where
\bea
\Pi_{\gamma \gamma'} (\omega, q) &=& \frac{1}{L^2} \sum_{\mathbf{p}}\frac{n_{\gamma}(\mathbf{p})-n_{\gamma'}(\mathbf{p+q})}{\omega-\varepsilon_{\gamma'}(\mathbf{p}+\mathbf{q})+\varepsilon_{\gamma}(\mathbf{p})+i\delta} \nonumber \\
& & \qquad \qquad |f_{\gamma \gamma'}(\mathbf{p},\mathbf{p+q})|^2. \label{eq:poldef}
\eea
where $n_{\gamma}(\mathbf{p})$ is the Ferm-Dirac distribution, $n_{\gamma}(\mathbf{p})=[\text{exp}\{\beta(\varepsilon_{\gamma}(\mathbf{p})-\mu)\}+1]$ with $\beta=1/k_{B}T$ and $\mu$ the chemical potential. The analytical expression for this polarization operator can be found in Ref. \onlinecite{hwang2007dielectric}.

In the long wavelength limit ($q \rightarrow 0$), the polarization operator of relativistic electrons is given by \cite{hwang2007dielectric}:
\begin{eqnarray}
\Pi_{\rm rel} (\omega, q)  \approx \left \{
\begin{array}{ll}  
\frac{D_0v_{F}^2q^2}{2\omega^2} \left [ 1-\frac{\omega^2}{4E_F^2}
\right ],  & v_{F} q < \omega < 2E_F  \\
D_0 \left [ 1 + i \frac{ \omega}{v_{F} q} \right ], & \omega <
v_{F} q,
\end{array} 
\right .
\label{pol_rel_q_small}
\end{eqnarray}
where $D_{0}= k_F /(2 \pi v_F)$ and $v_{F}$ is the 2D Fermi velocity and $E_{F}=v_{F} k_{F}$.

The contribution from the non-relativistic electrons is given by
\bea
\Pi_{\rm nr} (\omega, q) &=&  \sum_{\gamma= \uparrow,\downarrow}  \Pi_{\gamma\gamma} (\omega, q) \nonumber \\
&=& \frac{2}{L^2} \sum_{\mathbf{p}} \frac{n_{\rm nr}(\mathbf{p})-n_{\rm nr}(\mathbf{p+q})}{\omega-\varepsilon_{\rm nr}(\mathbf{p}+\mathbf{q})+\varepsilon_{\rm nr}(\mathbf{p})+i\delta}, \label{eq:nr}
\eea
where we have accounted for the spin degeneracy, and its analytical expression was first calculated by Stern \cite{PhysRevLett.18.546}. In the long wavelength limit this is given by \cite{giuliani2005quantum}:
\begin{eqnarray}
\Pi_{\rm nr} (\omega, q) &=& \frac{m}{2 \pi} \left(1 - \frac{\omega}{\sqrt{\omega^2- q^2 V_F^2}}\right) \nonumber \\
&\approx& \frac{m}{2 \pi} \frac{q^2 V_F^2}{\omega^2} \left[1+ \frac{3}{4} \frac{q^2 V_F^2}{\omega^2} \right],
\label{pol_non_rel_q_small}
\end{eqnarray}
where $V_F = \sqrt{2 (\mu-\Delta)/m}$.

The contribution that comes from the interaction between the two species is given by
\be
\Pi_{\rm int} (\omega, q) = \frac{2}{L^2} \sum_{\mathbf{p}} \sum_{\gamma, \gamma' = +, \uparrow,\downarrow}  \frac{n_{\gamma}(\mathbf{p})-n_{\gamma'}(\mathbf{p+q})}{\omega-\varepsilon_{\gamma'}(\mathbf{p}+\mathbf{q})+\varepsilon_{\gamma}(\mathbf{p})+i\delta} \, \frac{1}{2},
\ee
which we evaluate numerically. The factor of $1/2$ at the end comes from the overlap square of the wave functions. This integrand exhibits fast oscillatory behaviors and therefore, care must be exercised while evaluating it numerically. Here, we have neglected the interaction between the valence relativistic band and the parabolic band, the effect of which is negligible in the parameter regimes we are interested in.

In the following, we will consider two models of the interaction potential. In Model I, we consider a set-up where the two species are localized in two different planes. When these two planes coincide, the inter-species interaction is equal to the intra-species interactions, and when the two planes are infinitely apart, the inter-species interaction vanishes. The interaction potential is then given by
\be
V_{im}^{jn}(\mathbf{q}) = \frac{2\pi e^2}{\kappa \, q} \int_0^{\infty} dz_1 dz_2 \, \xi^{\ast}_{j}(z_1) \xi_{i}(z_1) \xi^{\ast}_{n}(z_2) \xi_{m}(z_2) e^{-q |z_1-z_2|},
\ee
where $\xi_i$'s are the transverse wave functions and $\kappa$ is the background material's dielectric constant. They are of the form
\be
\xi_{i} = \sqrt{\frac{2 \alpha_i \beta_i (\alpha_i + \beta_i)}{(\alpha_i-\beta_i)^2}} \left(e^{-\alpha_i z}- e^{-\beta_i z}\right),
\ee
as we are considering a semi-infinite material (on the $z$-direction) with the surface at $z=0$. The Dirac electrons are localized at $z = z_{\rm rel}\ll1$ and the non-relativistic electrons are localized at $z=z_{\rm nr} \ll 1$. We will fix the coefficients $\alpha_{\rm rel}$ and $\beta_{\rm rel}$ and vary the coefficients $\alpha_{\rm nr}$ and $\beta_{\rm nr}$ by varying the distance $d=|z_{\rm nr}-z_{\rm rel}|$. $\alpha_{\rm nr}$ and $\beta_{\rm nr}$ satisfy
\begin{widetext}
\bea
\frac{\log \tfrac{\alpha_{\rm nr}}{\beta_{\rm nr}}}{\alpha_{\rm nr}-\beta_{\rm nr}} &=& z_{\rm nr},  \\
\sqrt{\frac{2 \alpha_{\rm nr} \beta_{\rm nr} (\alpha_{\rm nr} + \beta_{\rm nr})}{(\alpha_{\rm nr}-\beta_{\rm nr})^2}} \left(e^{-\alpha_{\rm nr} z_{\rm nr}}- e^{-\beta_{\rm nr} z_{\rm nr}}\right) &=& \frac{\beta_{\rm rel}}{\alpha_{\rm rel}} \sqrt{\frac{2 \alpha_{\rm rel} \beta_{\rm rel} (\alpha_{\rm rel} + \beta_{\rm rel})}{(\alpha_{\rm rel}-\beta_{\rm rel})^2}}  \left(e^{-\tfrac{\alpha_{\rm rel}}{\alpha_{\rm rel}-\beta_{\rm rel}}}- e^{-\tfrac{\beta_{\rm rel}}{\alpha_{\rm rel}-\beta_{\rm rel}}}\right), \nonumber
\eea
\end{widetext}
such that as $z_{\rm nr} \rightarrow z_{\rm rel}$, $\xi_{\rm nr} \rightarrow \xi{\rm rel}$. We find that in general, there is only one (undampened) plasmon mode when the planes in which the two species are localized are nearby (or when the difference between the inter-species and intra-species interactions are negligible). 

To study the effects of the inter-species interaction in more details, we consider a less realistic set-up in Model II, where the interaction is given by 
\be
V_{im}^{jn}(\mathbf{q}) = \frac{2\pi e^2}{\kappa \, q}, \ {\rm if} \ i=j=m=n,
\ee
and 
\be
V_{im}^{jn}(\mathbf{q}) = C\ \frac{2\pi e^2}{\kappa \, q}, 
\ee
otherwise. Here, $C \in [0,1]$, where $C=0$ corresponds to the case where the inter-species interaction vanishes and $C=1$ corresponds to the case where the inter-species interaction is of the same strength as the intra-species interaction. The results obtained using this model qualitatively agree with those of Model I. The Model II is convenient for studying the effects of the inter-species interaction because one can obtain relatively accurate solutions to Eq. (\ref{eq:det}) for small but finite values of momenta without using a significant amount of computation time.

\section{Results \label{res}}

\begin{figure}[t]
\begin{center} 
\includegraphics[width=3 in]{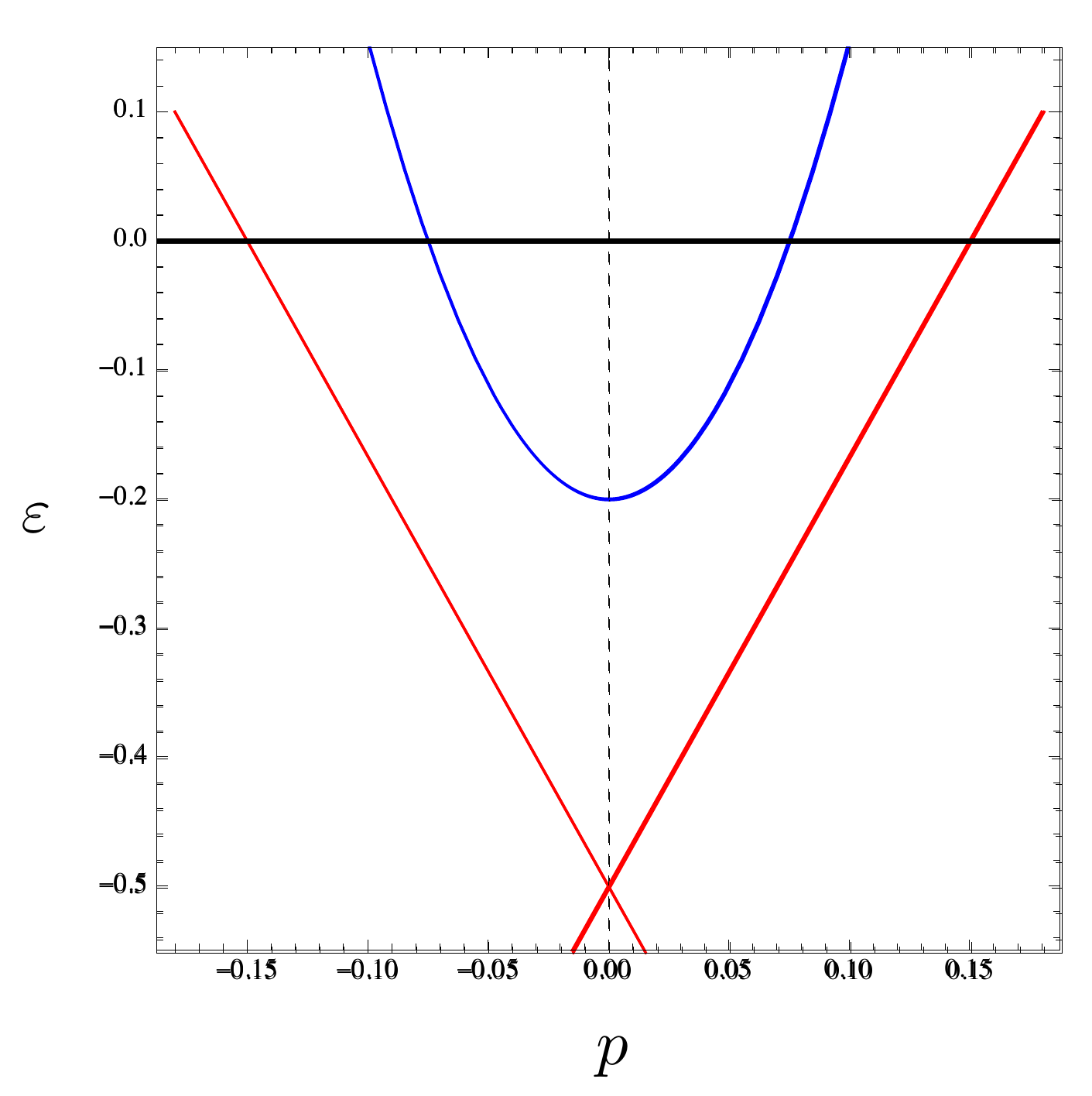}
\end{center}  
\caption{Dispersion relations for the surface states (red) and the 2D non-relativistic electron gas (blue). Parameters are chosen in accordance with the experiment of Ref. \onlinecite{bianchi2010coexistence}. \label{fig:states}}
\end{figure}

\begin{figure*}[t]
\begin{center} 
\includegraphics[width=7 in]{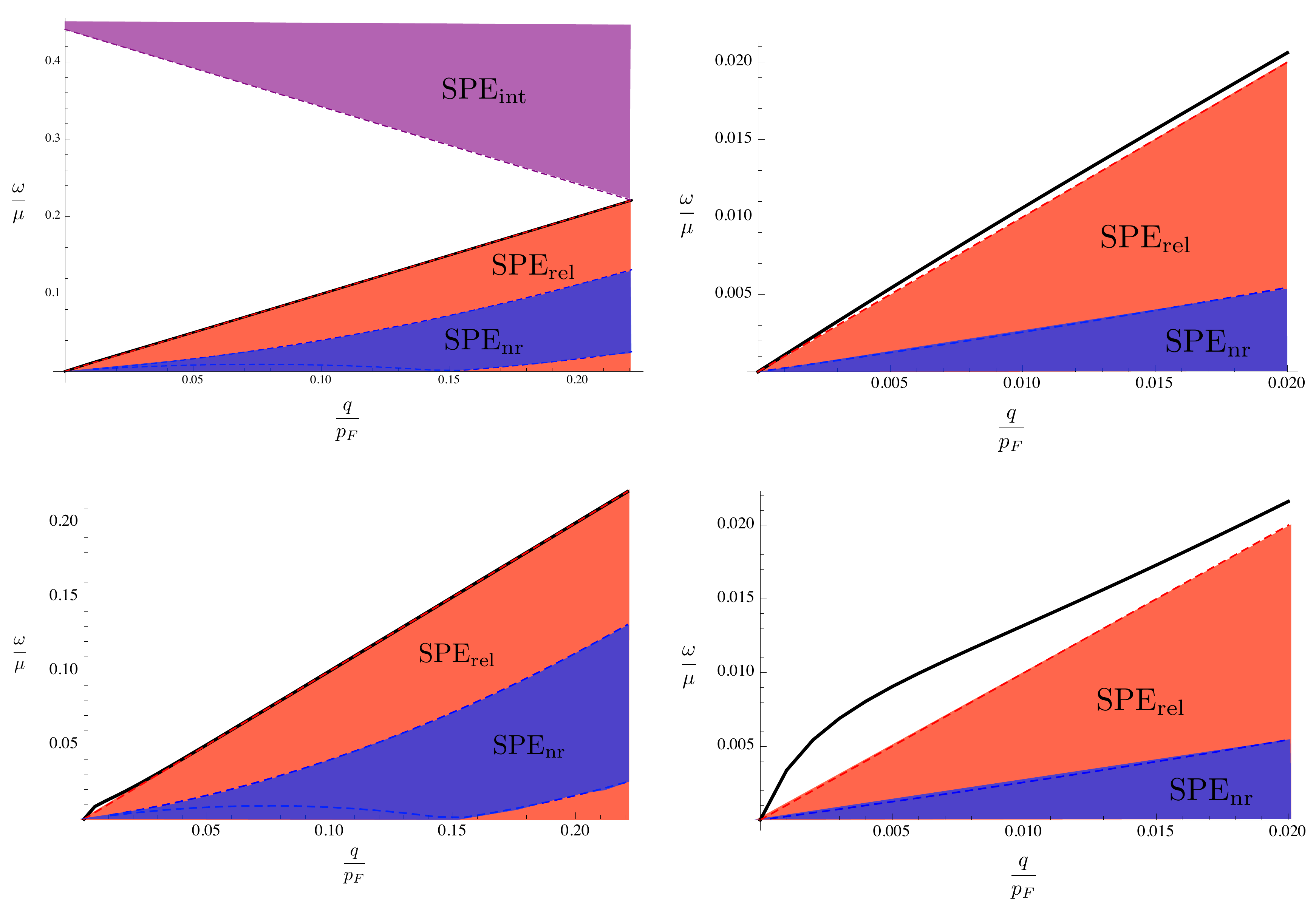}
\end{center}  
\caption{Plasmon spectrum obtained using Model I along with the SPE regimes for $d=0$ (top) and $d= 500 \mathring{\rm A}$ (bottom). The plasmon is plotted as the black line, while SPE$_{\rm rel, \ nr, \ int}$ are shaded in red, blue and purple, respectively.  \label{fig:ILM}}
\end{figure*}

For concreteness, we will take parameters relevant to Bi$_2$Se$_3$: $v =10/3$ eV$\mathring{\rm A}$, $m = 0.014$  $\mathring{\rm A}^{-2}/$eV, $\Delta = 0.3$ eV and $\mu = 0.5$ eV \onlinecite{bianchi2010coexistence}, the dimensionless fine structure constant is $r_s = 0.09$ \onlinecite{efimkin2012collective} and the parameters for the transverse wave function are $\alpha_{\rm rel}=3.76 \times 10^{-3} \mathring{\rm A}^{-1}$  and $\beta_{\rm rel}= 3.89 \mathring{\rm A}^{-1}$ \onlinecite{PhysRevB.82.045122}. The dispersion relations are plotted in Fig. \ref{fig:states}. Furthermore, by analyzing the imaginary part of the polarization operator, we find the single particle excitation (SPE) continuum. The region where $\Im[\Pi_{\rm rel}] \ne 0$ is given by $\omega/\mu - q/p_F<0$ and is denoted by SPE$_{\rm rel}$, the region where $\Im[\Pi_{\rm nr}] \ne 0$ is given by $-1 < \omega/(q V_F) - q V_F/(2 (\mu- \Delta))<1$ and is denoted by SPE$_{\rm nr}$ and the region where  $\Im[\Pi_{\rm int}] \ne 0$ is given by  $\omega/\mu + q/p_F > 0.442$, which is denoted by SPE$_{\rm int}$.

The results for Model I are shown in Fig. \ref{fig:ILM}. When the two species are localized on the same plane, we find only one plasmon mode, described by a linear dispersion relation
\be
\frac{\omega}{\mu} \approx 
1.005 \, \frac{q}{p_F}.
\ee
This mode lies outside SPE$_{\rm rel}$ but is very close to its boundary. Furthermore, as we increase the distance $d$ between the planes on which each of the species reside, the plasmon dispersion at small momenta deviates from the equation above. In order to study the small momentum behavior and to understand the reason why there is only one plasmon mode, we now turn to Model II.  

\begin{figure*}[t]
\begin{center} 
\includegraphics[width=7 in]{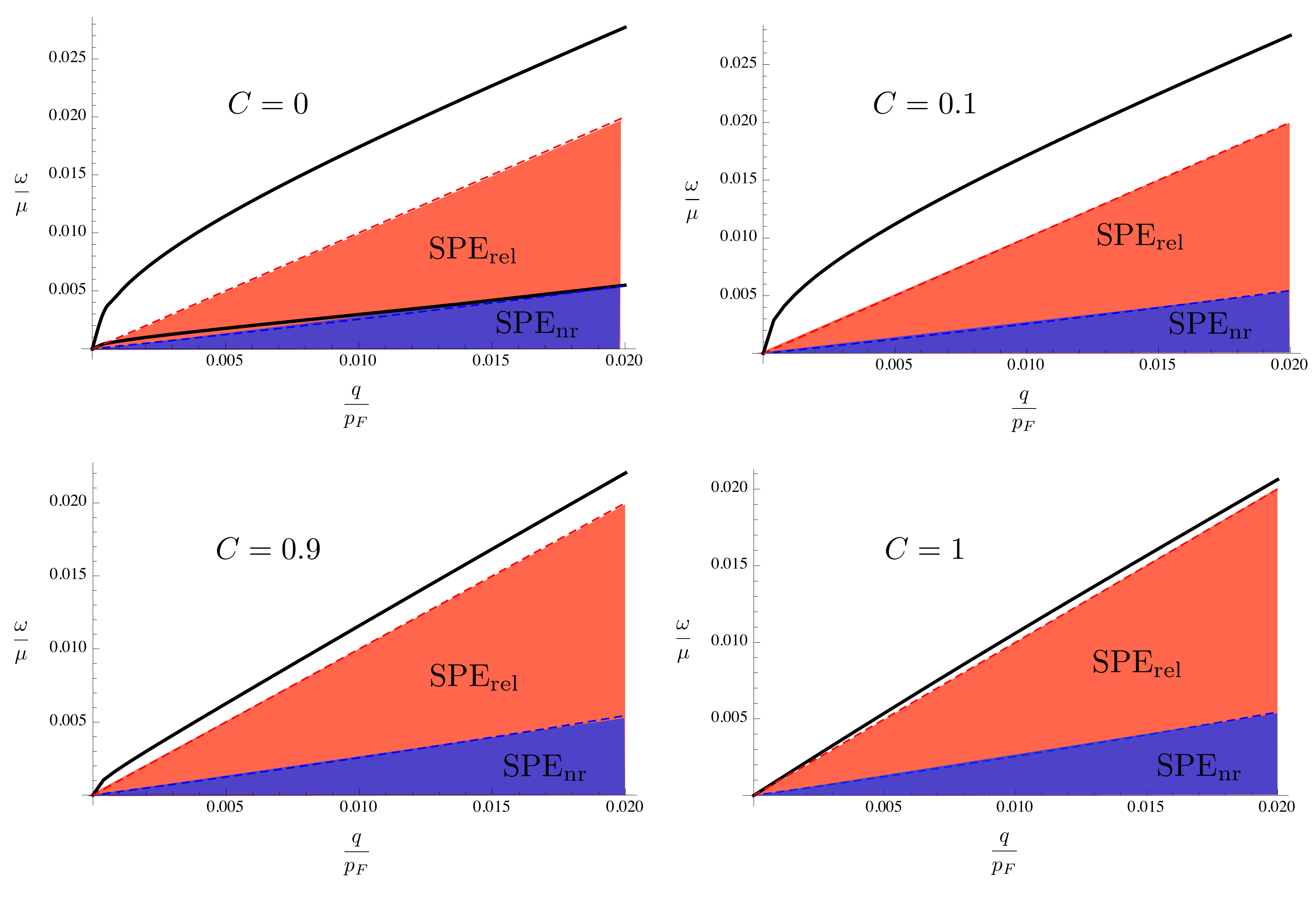}
\end{center}  
\caption{Plasmon spectrum obtained using Model II for different values of $C$.  \label{fig:IILM}}
\end{figure*}

\begin{figure}[t]
\begin{center} 
\includegraphics[width=3.5 in]{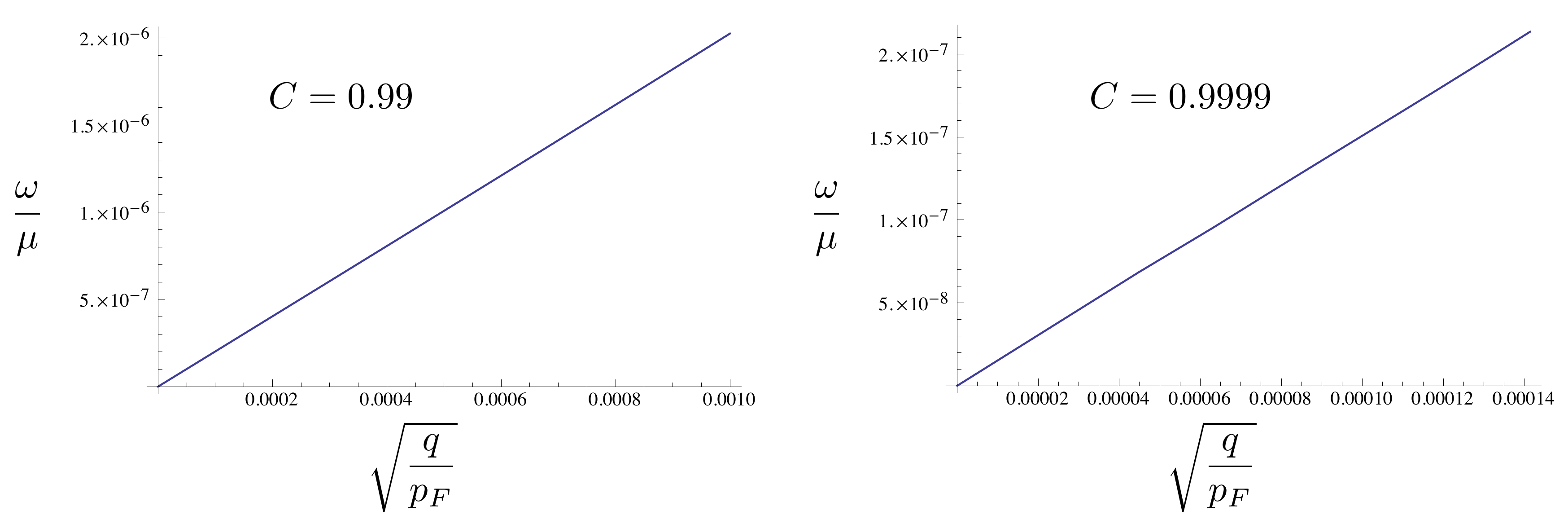}
\end{center}  
\caption{Plasmon spectrum as a function of $\sqrt{q}$ for small $q$.  \label{fig:sqrt}}
\end{figure}

The results of Model II are plotted in Fig. \ref{fig:IILM}. When $C=0$, there are two plasmon modes, both exhibiting $\omega \sim \sqrt{q}$ behavior at small $q$. The plasmon of the Dirac electrons lies outside the SPE regime, but the plasmon of the non-relativistic electron is mostly buried underneath SPE$_{\rm rel}$. Therefore, even for vanishing inter-species interaction, there is only one undamped plasmon mode. Moreover, as one turns on the inter-species interaction, the damped plasmon mode disappears. 

Interestingly, the dispersion relation $\omega \sim \sqrt{q}$ at small $q$ persists even as one approaches $C\rightarrow 1$. As can be seen in Fig. \ref{fig:sqrt}, when $C$ is close but not equal to unity, this behavior is visible up to $\omega \sim 10^{-3} \sqrt{1-C}$, and only when $C=1$ do we observe a linear plasmon spectrum at small $q$ (within our numerical accuracy). In the light of the experimental result of Ref. \onlinecite{PhysRevB.85.085443}, where $\omega \sim \sqrt{q}$ at small $q$, this suggests that in Bi$_2$Se$_3$, the strength of the inter-species interaction differs from that of the intra-species interaction. This is presumably due to the two species being localized on different planes near the surface.

\begin{figure*}[t]
\begin{center} 
\includegraphics[width=7 in]{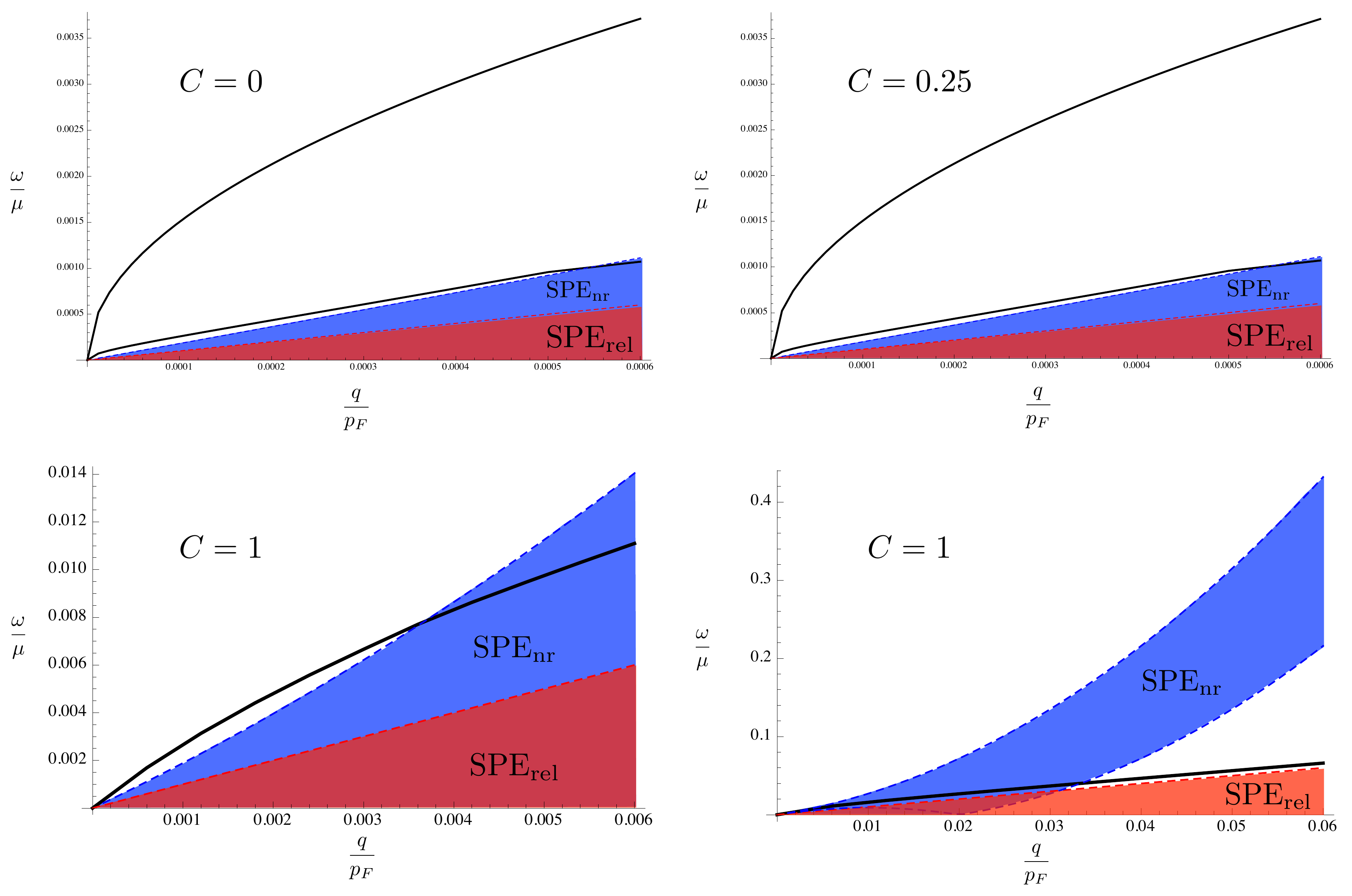}
\end{center}  
\caption{Plasmon spectrum obtained using Model II for small mass.  \label{fig:IISM}}
\end{figure*}

In the above discussion, we have assumed parameters relevant to Bi$_2$Se$_3$. We now ask whether two distinct modes can be seen in some parameter regime. In the above, we had a single undamped plasmon mode even when the inter-species interaction was switched off ($C=0$). It is natural then to ask what happens in the case of two undamped plasmon modes at $C=0$. 

\begin{figure}[h]
\begin{center} 
\includegraphics[width=3.5 in]{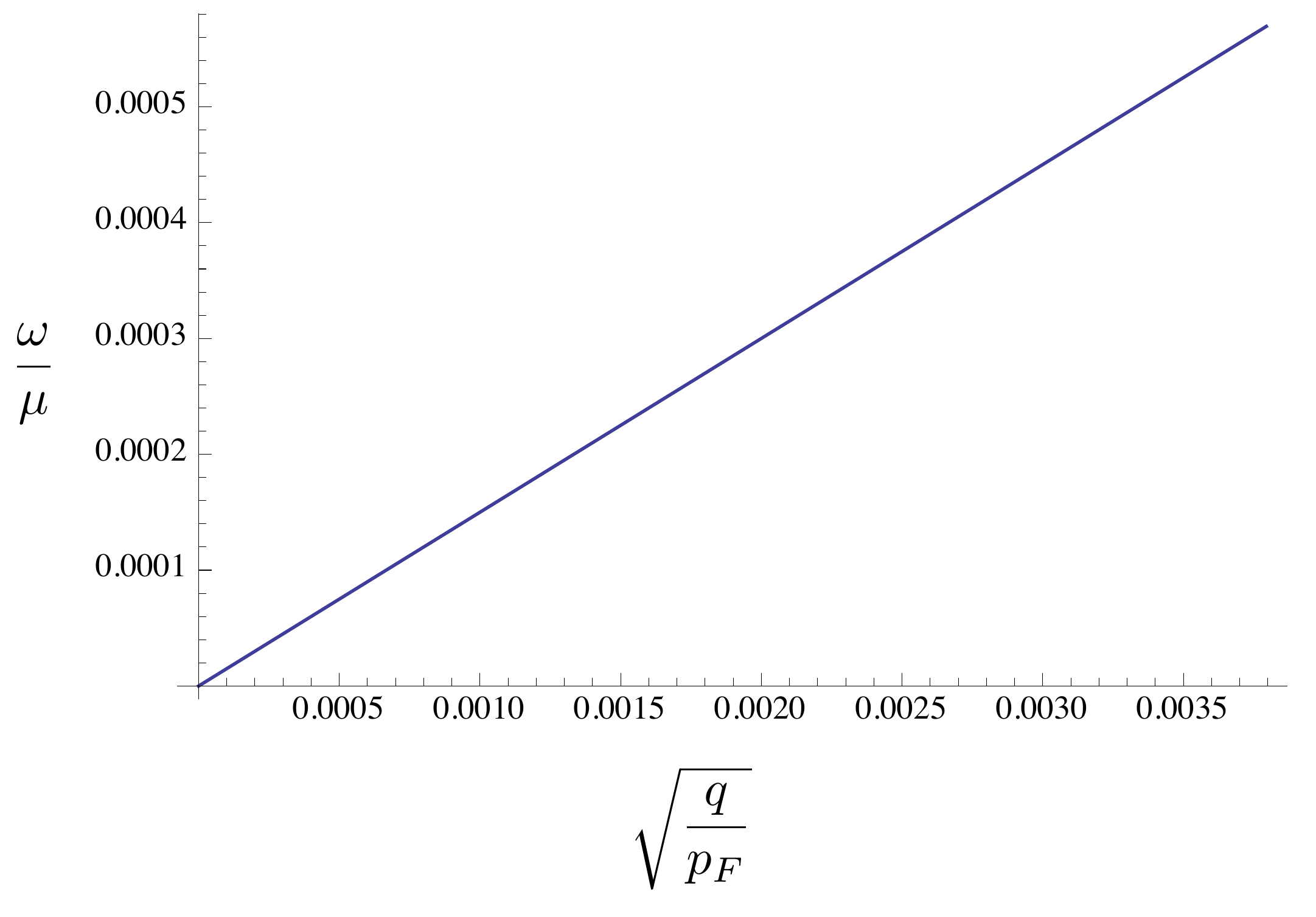}
\end{center}  
\caption{$\omega \sim \sqrt{q}$ behavior of the dispersion relation at $C=1$ for small mass.  \label{fig:sqrt1}}
\end{figure}

We find that in a broad range of parameter regime, there is only a single undamped plasmon mode at $C=0$. Nevertheless, by tuning the mass of the non-relativistic electrons small enough, it is possible to obtain two undamped plasmon modes at $C=0$. We will consider now such case where we take $m = 2.5 \times 10^{-4}$  $\mathring{\rm A}^{-2}/$eV while keeping the rest of the parameters the same as in the previous case. The results for Model II are shown in Fig. \ref{fig:IISM}.

Here, SPE$_{\rm int}$ is well beyond our energy range. This is expected because the energy required to excite a particle from the filled relativistic band onto the parabolic band is higher than the previous case. We find that at small but non-vanishing values of $C$, there are two undamped plasmon modes. As the inter-species interaction increases, the mode with lower energy gets pushed below the boundary of SPE$_{\rm nr}$ and finally disappears. Therefore, at larger values of $C$, we are again left with one plasmon mode. Interestingly, this mode retains the scaling behavior $\omega \sim \sqrt{q}$ even when $C=1$ (see Fig. \ref{fig:sqrt1}).

For Model I, for reasonable distance $d$ (up to about 50 nm), we only find a single plasmon mode with similar qualitative behaviors as the results for Model II.

\section{Summary and Discussions \label{summ}}

In this article, we studied the collective excitations of a two-dimensional electron system consisting of two species of charge carriers: relativistic and non-relativistic electrons. We considered a model where the relativistic and non-relativistic electrons are localized in two spatially separated planes; the separation between them can be varied to tune the relative strengths of the inter- and the intra-species interaction. In order to study the effects of the relative strengths of the inter-species and intra-species interactions on the behaviors of the plasmon modes, we also considered a simplified model in which the relative strength between the interactions are independent of the exchange momentum.

As expected, when the inter-species interaction vanishes, we have two plasmon modes. However, in a broad range of parameter regime, only one of these plasmon modes resides outside the SPE regime and thus survives Landau damping. As for the undamped plasmon mode, its dispersion relation exhibits $\omega \sim \sqrt{q}$ at small $q$. For the parameters relevant to Bi$_2$Se$_3$, the dispersion relation becomes linear when the strength of the inter-species interaction becomes equal to the strength of the intra-species interaction. In the light of the experimental result of Ref. \onlinecite{PhysRevB.85.085443}, this suggests that the inter-species interaction is not equal to the intra-species interaction. This is presumably due to the Dirac surface states and non-relativistic electrons being localized in two non-coincidental planes.

For small enough non-relativistic electron mass, it is possible to obtain two undamped plasmon modes at vanishing inter-species interaction. As the inter-species interaction increases, one of these modes enters the SPE regime, gets damped and thereby disappears.

In this article, we have considered a semi-infinite TI where we have neglected the interaction with states at the opposite surface. However, in thin film TIs such as those used in the experiment of Ref. \onlinecite{PhysRevB.85.085443}, the interaction between states from both surfaces cannot be neglected. 

\section*{Acknowledgement}
This work is supported in part by NSF grant DMR-0820404 Penn State MRSEC (ACB and JH), and by DOE grant no. DE-SC0005042 (JKJ). We thank Dr. A. Principi for bringing to our attention Ref.~\onlinecite{PhysRevB.86.085421}.

\bibliography{References}

\end{document}